\documentclass[twocolumn,prl,amsmath,amssymb,showpacs,superscriptaddress,floatfix]{revtex4}
\usepackage{graphicx}
\usepackage{bm}
\usepackage{lineno,hyperref}
\usepackage{epstopdf}
\usepackage{epsf}
\begin{document}
\title{Anomalous behavior of dispersion of longitudinal and transverse collective excitations in water}

\author{Yu. D. Fomin, E. N. Tsiok, V. N. Ryzhov, and V. V. Brazhkin}
\affiliation{ Institute for High Pressure Physics RAS, 108840
Kaluzhskoe shosse, 14, Troitsk, Moscow, Russia}

\date{\today}

\begin{abstract}
We study the dependence of the excitation frequency of water along an isochore and an isotherm crossing
the region of density anomaly. We have shown that the frequency of the longitudinal excitations demonstrated
anomalous dependence on temperature along the isochore. At the same time the dependence for both
longitudinal and transverse excitation frequencies on density along the isotherm are very modest or even
negligible in rather wide range of densities. This kind of behavior also seems anomalous in comparison with the ordinary liquids.
\end{abstract}


\maketitle

\section{Introduction}

Collective excitations are of great importance in the physics of solids \cite{solidstate,kittel}.
Many properties of solids may be efficiently described on a basis of the collective excitations,
i.e. phonons. Only recently it has been recognized that collective excitations are of the same
importance for liquids \cite{boon,ruocco}. In particular, a theory of heat capacity of liquids
was proposed in Ref. \cite{phonon-liq}.

Altough it was widely known for a long time that viscous liquids can demonstrate phonon-like excitations
similar to the ones in crystals, detailed investigation of collective excitations in liquids
started only several decades ago. The reason for this is that experimental study of longitudinal
and in particular transverse waves in liquids requires very high quality x-ray or neutron
installations which became available only recently. Currently, high quality experimental data
are available for several liquid metals \cite{hos-fe-cu-zn,psd-ga,ga-mokshin,psd-fe,psd-sn,psd-fe-1,psd-sn-1},
ionic solutions \cite{nacl,kcl,nai}, water \cite{wat-mon,wat-jap} and some more substances.

Several theoretical models were proposed to describe the collective excitation in liquids.
Good agreement with experimental data was obtained in the model introduced in Ref. \cite{mokshin1}
(see also \cite{mokshin2,mokshin3,mokshin4,mokshin5,mokshin6} for examples of successful implementation
of this model).

The collective excitations in liquids were also widely studied by means of computer simulation methods.
The first attack on the system was made in classical set of papers by D. Levesque, L. Verlet \cite{lj4}.
After that numerous papers appear in the literature where collective excitations in different liquids
were studied by computer simulation methods.


It is well known that some liquids demonstrate unusual properties usually called anomalies. The most
well known anomalous liquid is water which demonstrates more then 70 anomalies, such as density, diffusion, structural
anomalies and many others \cite{wateranom}. In particular, there are some anomalies of the thermodynamic
properties, such as density anomaly, i.e. negative thermal expansion coefficient. In solids it is well
known that in the case of negative thermal expansion coefficients there are some phonon
branches which demonstrate unusual density dependence: while in normal crystal the phonon
frequency increases with density, in the case of negative thermal expansion some
phonon branches have the frequency which decreases with density \cite{nte-review}. Because of this
one can expect that the frequency of collective excitations of liquids with density anomaly
also demonstrates unusual behavior.

This assumption was confirmed in our previous work where we studied the dispersion curves
of longitudinal excitation frequency along isochors which cross the region of the density anomaly.
It was found that while in normal liquid the frequency increases with temperature, the excitation
frequency of the liquid with density anomaly can decrease. In the present paper we extend these
findings. We perform simulation of dispersion curves of water along isochors and isotherms
crossing the region of the density anomaly and monitor the temperature and density
dependence of the excitation frequency and show that in both cases the behavior of the
frequency is anomalous.

\section{System and Methods}

In the present paper we simulate water in the an SPC/E model \cite{spce}. Although this model
is not very accurate \cite{vega-comparison,vega-comparison1}, it gives correct qualitative
description of the main anomalous features of water. At the same time SPC/E model is computationally
cheaper then more accurate ones, for instance, TIP4P/2005. Because of this we chose to
study the SPC/E water for rapid evaluation of qualitative features of the behavior of water.

The temperature of maximum density (TMD) of the SPC/E water at atmospheric pressure is 241 K \cite{vega-comparison1}.
Because of this we simulate water along an isotherm $T=240$ K in order to cross the region
of density anomaly. The density is varied from $\rho_{min}=0.95$ $g/cm^3$ up to $\rho_{max}=1.05$ $g/cm^3$.
Another set of simulations is along an isochor $\rho=1.01$ $g/cm^3$. The temperature is changed from
$T_{min}=220$ K up to $T_{max}=700$ K.

In all cases we performed molecular dynamics simulations of 4000 water molecules in a cubic box with periodic
boundary conditions. The initial structure is a high temperature structure at the given density. The system
was equilibrated for $1 \cdot 10^7$ steps with the time step $dt=1$ fs. After that the system was simulated
for more $5 \cdot 10^7$ steps with $dt=0.1$ fs for calculating the averages. This interval was divided into
10 blocks and then the results of all blocks were averaged.

In order to find the excitation frequencies of water we calculated the correlation functions of fluxes of the velocity current.
The longitudinal and transverse parts of these correlation functions are defined as
\begin{equation}
C_L(k,t)=\frac{k^2}{N}\langle J_z({\bf k},t) \cdot J_z(-{\bf
k},0)\rangle
\end{equation}
and

\begin{equation}
C_T(k,t)=\frac{k^2}{2N} \langle J_x({\bf k},t)\cdot J_x(-{\bf
k},0)+J_y({\bf k},t) \cdot J_y(-{\bf k},0)\rangle
\end{equation}
where $J({\bf k},t)=\sum_{j=1}^N {\bf v}_j e^{-i{\bf k r}_j(t)}$
is the velocity current and the wave vector $\bf{k}$ is directed
along the z axis \cite{hansenmcd,rap}. The excitation frequencies are
calculated as the location of the peak of the
Fourier transform of these functions. The same methodology was successfully employed to investigation
of collective excitations in many different systems, for instance, liquid metals (see, for instance, Refs.
\cite{mokshin1,mokshin2,mokshin3,mokshin4,mokshin5,mokshin6,lithium}), supercritical metals \cite{hg},
binary mixtures \cite{binary}, molecular liquids \cite{kostya}, and many other systems.

In the present work we calculate the dispersion curves of the SPC/E water along isochore $\rho=1.01$ $g/cm^3$
and isotherm $T=240$ K and monitor the evolution of the excitation frequency with temperature and the density
respectively. In the case of simple liquids the frequency increases both under isochoric heating and
isothermal compression. As it was shown in our previous publication \cite{sp-anom} the anomalous liquids can demonstrate
anomalous decreasing of the frequency when the temperature is increased at constant density. By analogy with
crystals with negative thermal coefficient \cite{nte-review} we may expect that anomalous behavior of frequency of transverse
excitations can take place in water in the region of density anomaly.

All simulations were performed in LAMMPS simulation package \cite{lammps}.

\section{Results and Discussion}

\begin{figure}
\includegraphics[width=8cm]{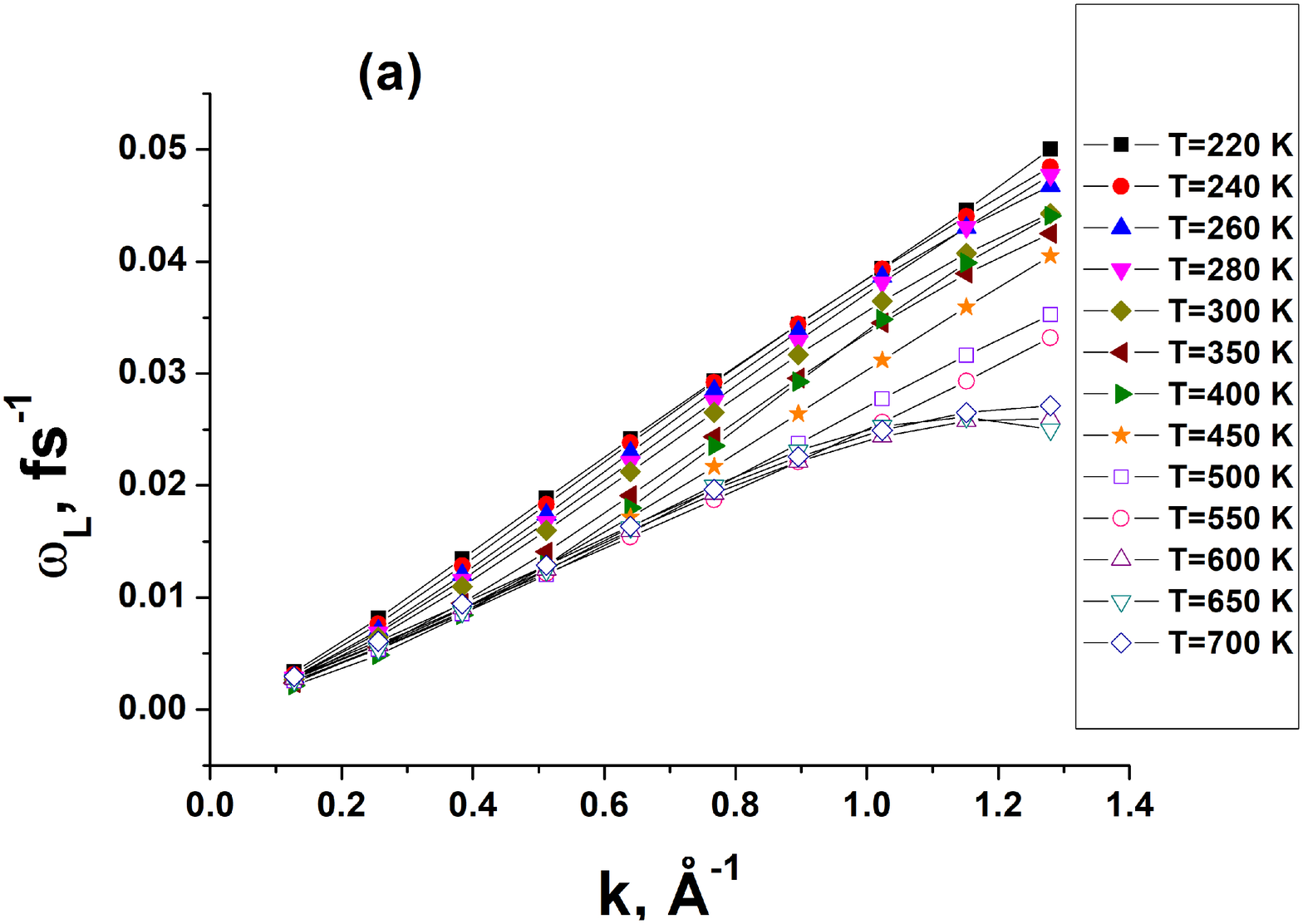}%

\includegraphics[width=8cm]{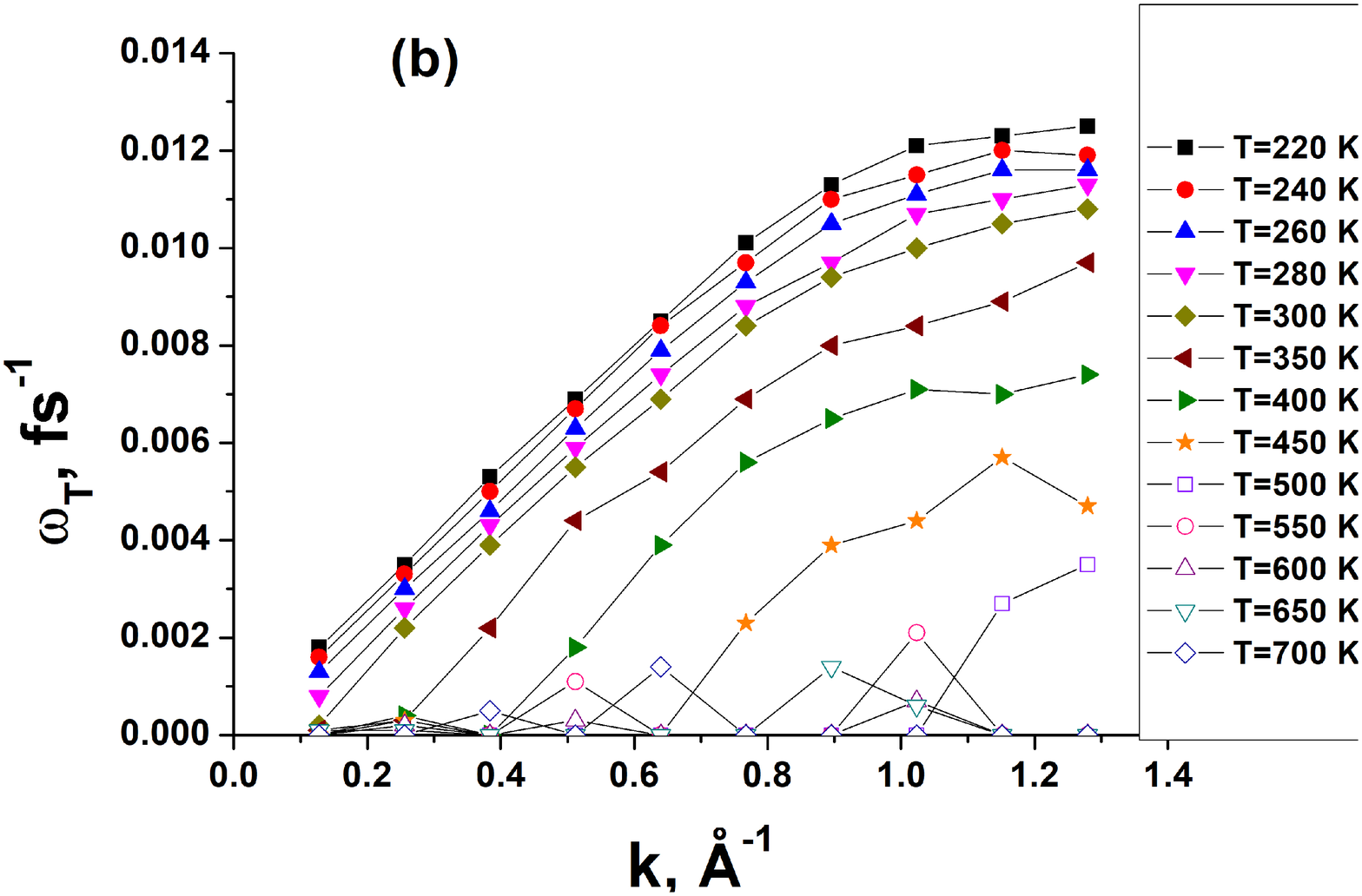}%

\caption{\label{r101} Dispersion curves of the SPC/E water along isochore $\rho=1.01$ $g/cm^3$. Panel (a)
shows the dispersion of the longitudinal excitations, while panel (b) the transverse excitations.}
\end{figure}

We start the discussion from the dispersion curves at $\rho=1.01$ $g/cm^3$ and different temperatures.
Fig. \ref{r101} shows the dispersion curves of longitudinal (panel a) and transverse (panel b) excitations.
One can see that as the temperature increases the curves of longitudinal excitations go downward, i.e. the frequency decreases.
Therefore, the system does demonstrate the anomalous dependence of the excitation frequency. Only at the temperature as
high as 700 K the curves turn to increasing of the frequency with temperature.

The frequency of transverse excitations also decrease with temperature. However, for the transverse waves this is
normal behavior. The transverse excitations take place in liquids close to the melting line. When the temperature
increases they disappear at the Frenkel line (FL) \cite{frpre,frprl,ufn}. The Frenkel line for the SPC/E model of water was calculated in Ref. \cite{fr-water}
(see also \cite{kostya-ch} for the Frenkel line of the TIP4P/2005 model of water).
For $\rho=1.01$ $g/cm^3$ the temperature of the FL is about 500 K. From Fig. \ref{r101} (b) one can see that
only two points close to the boundary of the Brillouin zone remain while at $T=550$ K no transverse waves are observed.
Therefore, our study is consistent with previous calculations of the FL of the SPC/E water.

\begin{figure}
\includegraphics[width=8cm]{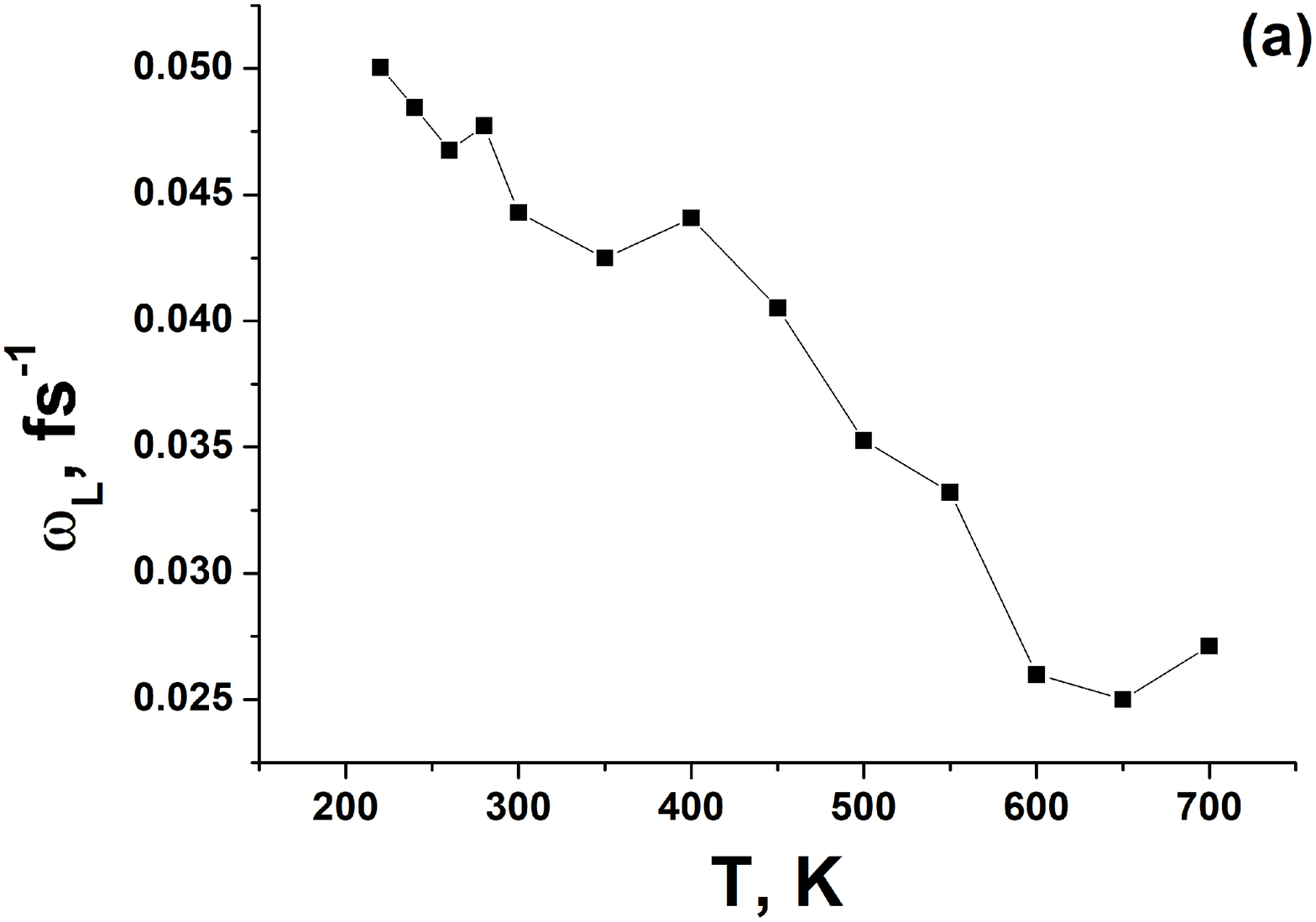}%

\includegraphics[width=8cm]{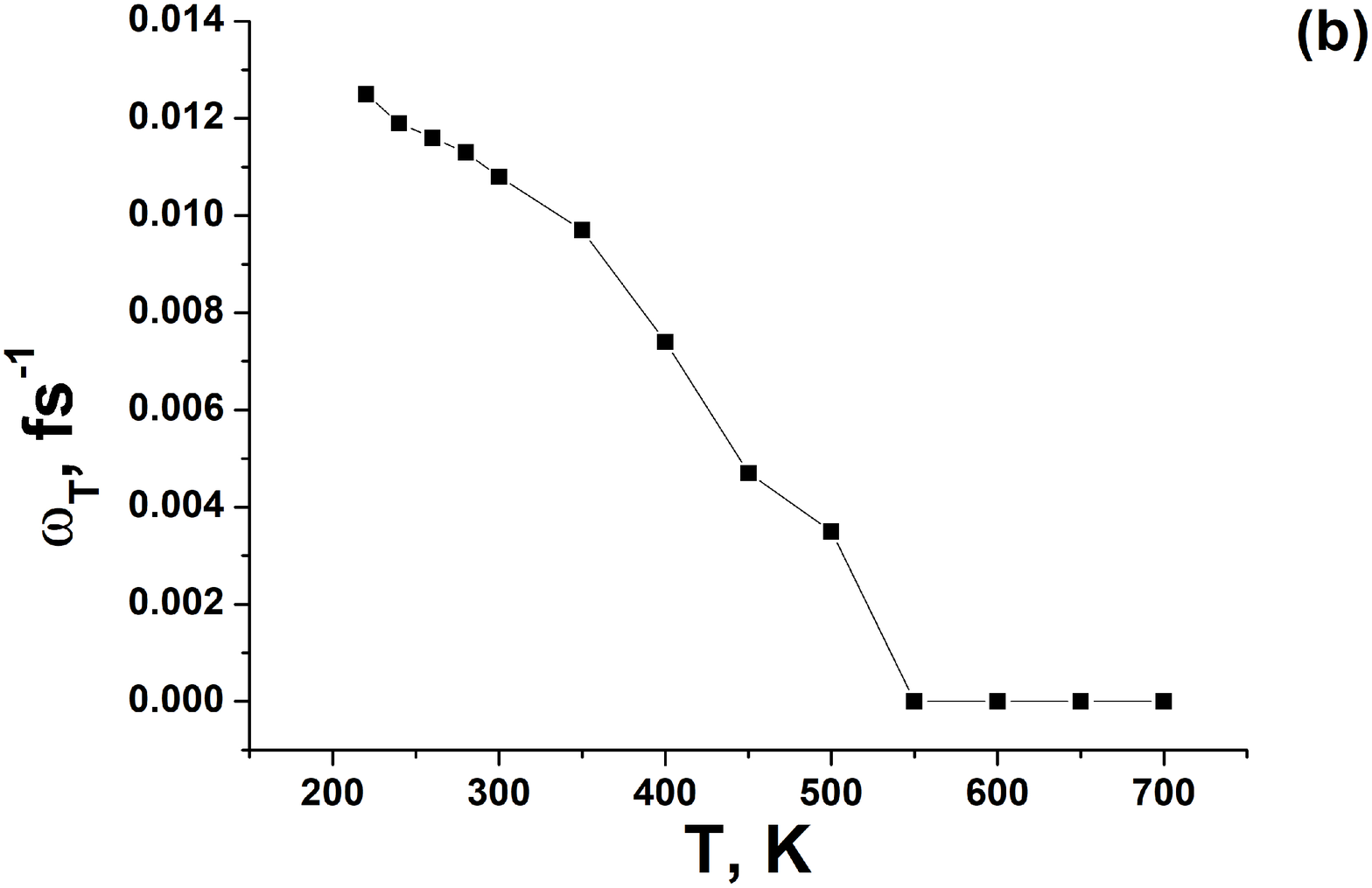}%

\caption{\label{wlwtr101} The temperature dependence of the excitation frequency at fixed wave vector $k=1.28 \AA^{-1}$ along $\rho=1.01$ $g/cm^3$. (a)
frequency of the longitudinal excitations, (b) the transverse excitations.}
\end{figure}

In order to make the effect clearer we show the temperature dependence of the excitation frequency at constant
magnitude of the wave vector. Fig. \ref{wlwtr101} shows the temperature dependence of the excitation frequency of longitudinal and
transverse excitations at $k= 1.28 \AA^{-1}$ which is the largest wave vector in our study. This value of k is selected
since it is the closest one to the boundary of the Brillouin zone. In spite of noisy behavior the tendency of the
longitudinal frequency to decrease with temperature is apparent. Only at temperatures as high as 650-700 K
the frequency starts to increase again.

In the case of the transverse excitations the situation is typical for a liquid. Liquids demonstrate some
shear rigidity at low temperature. However, the rigidity decreases with temperature, and therefore
the transverse excitations become depresses and finally disappear. This is exactly what we observe in the
present case.

\begin{figure}
\includegraphics[width=8cm]{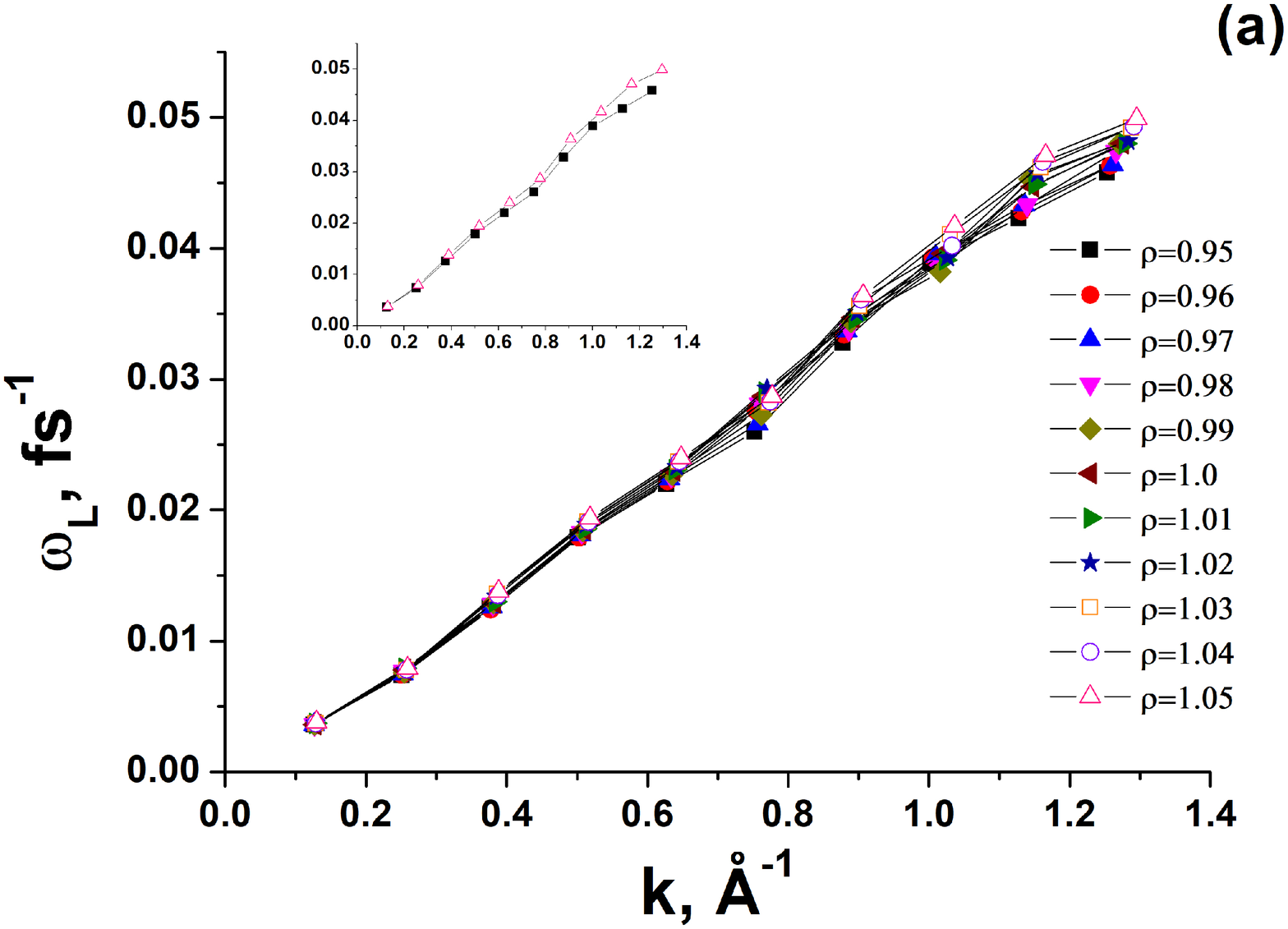}%

\includegraphics[width=8cm]{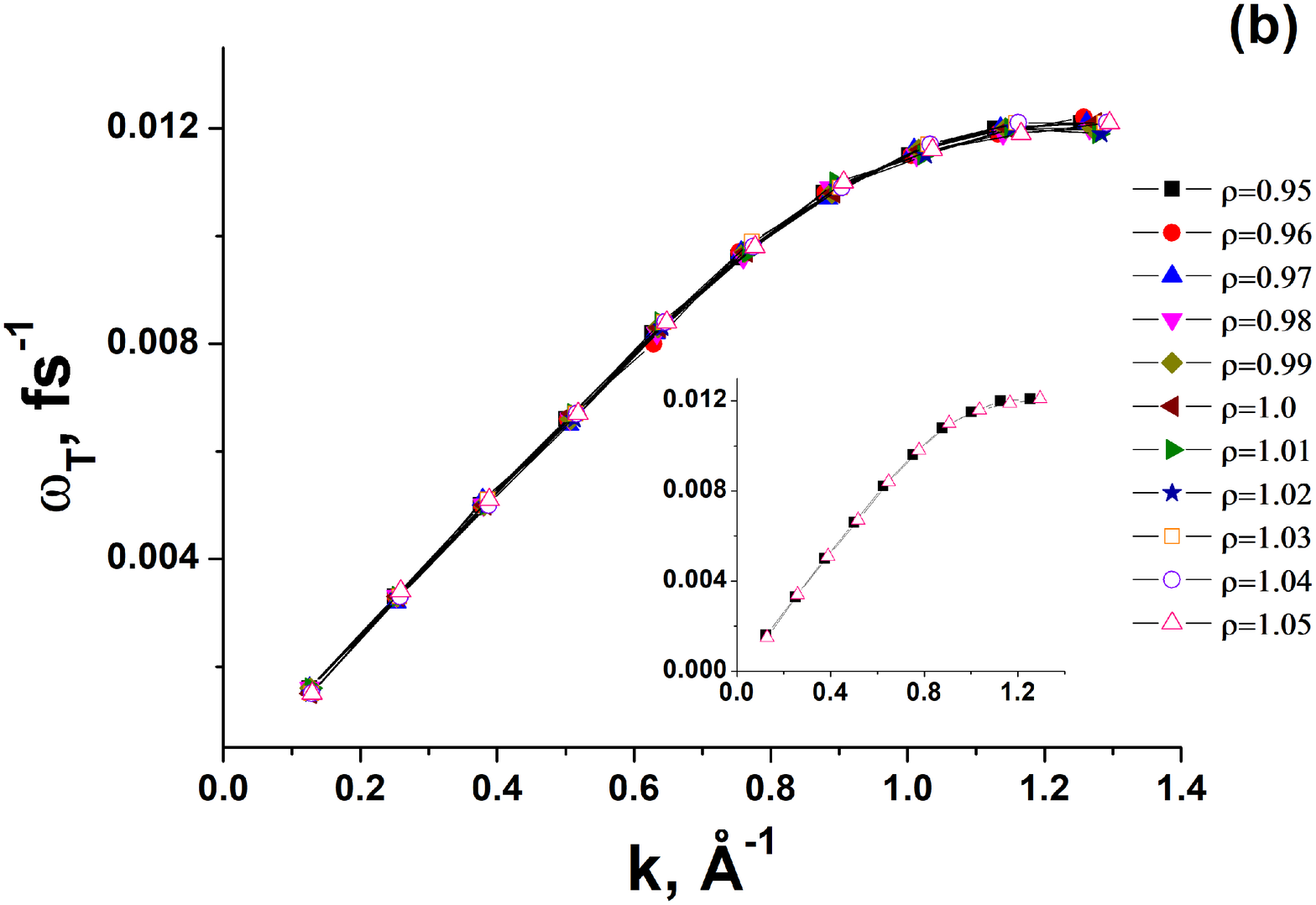}%

\caption{\label{wlwtt240} The dispersion curves of the SPC/E water along the isotherm $T=240$ K. Panel (a)
shows the dispersion of the longitudinal excitations, while panel (b) the transverse excitations. The insets on
both panels enlarge the lowest and the highest density ($\rho=0.95$ and $1.05$ $g/cm^3$). The
numbers on the plots mean the density in $g/cm^3$.}
\end{figure}

Consider the dispersion curves at isotherm $T=240$ K and different densities (Fig. \ref{wlwtt240}). We see that
the effect of the density on the frequency of longitudinal excitations is very modest. From the inset of the
panel (a) we see that the frequencies of the last points of the curves at $\rho=0.95$ $g/cm^3$ and
$1.05$ $g/cm^3$ are $\omega_{0.95}=0.0458 fs^{-1}$ and $\omega_{1.05}=0.05 fs^{-1}$, i.e. the change is about
$8$ percent.

In the case of transverse excitations the curves fall on the same master curve, i.e. no dependence of
frequency on the density is indicated.

The results for both longitudinal and transverse excitations are in contrast to the behavior of simple
liquids. In our previous works we studied the dispersion curves in the Lennard-Jones \cite{welj} and soft sphere systems \cite{wesoft}.
It was shown that both longitudinal and transverse branches strongly depend on the density. 

The behavior of water is remarkably more complex then the one of a simple liquid. Among the most common interpretations
of the water behavior is the existence of the second critical point in deeply supercooled region \cite{water-llpt}. 
However, many aspects of the behavior of water can be studied within the framework of some simplified systems
which allow clear interpretation of the results. One of such systems was introduced
and widely studied in our previous works \cite{s135,s135a,s135b,s135c,s135d,s135e}. It was shown that this system
demonstrates complex phase diagram and water-like anomalies, for instance, density and diffusion ones. 
This system is characterized by a multiscale potential which induces quasibinary behavior of the system
which can be clearly seen from evolution of the radial distribution functions of the system
along isochores or isotherms \cite{s135}. Such behavior may be interpreted as a smooth structural crossover
between the structures with different length scales of the potential. Such crossover is responsible
for many anomalous properties like density, diffusion and structural anomalies or high value of 
the heat capacity \cite{cvlarge,cvlarge1}. In the context of this work the smooth structural crossover
makes the system effectively softer under isochoric heating which make the frequencies smaller. At the same time
while the frequency should increase with density along isotherm opposite effect takes place in the anomalous region
which balances the rise of the frequency. Similar behavior should be expected in other anomalous liquids, for instance, 
liquid silicon \cite{sastry} whose qualitative behavior is very close to the one of the model system from \cite{s135}.


\section{Conclusions}

We study the dependence of the excitation frequency of water along the isochore and the isotherm crossing
the region of density anomaly. We show that the frequency of the longitudinal excitations demonstrate
anomalous dependence on temperature along the isochore. At the same time the dependence of both
longitudinal and transverse excitation frequencies on density along isotherm are very modest or even
negligible, which can also be considered as anomalous comparing to the behavior of these quantities to the one
in a simple liquid.


This work was carried out using computing resources of the federal
collective usage center "Complex for simulation and data
processing for mega-science facilities" at NRC "Kurchatov
Institute", http://ckp.nrcki.ru, and supercomputers at Joint
Supercomputer Center of the Russian Academy of Sciences (JSCC
RAS). The work was supported by the Russian Foundation of Basic Research (Grants No 18-02-00981).

\end{document}